# Modeling of Nuclear Waste Forms: State-of-the-Art and Perspectives

**Piotr. M. Kowalski[1,2,*], Steve Lange[1,2], Guido Deissmann[1,2], Mengli Sun[1,2,3], Kristina O. Kvashnina[4,5], Robert Baker[6], Philip Kegler[1,2], Gabriel Murphy[1,2], Dirk Bosbach[1,2]**

[1] Institute of Energy and Climate Research (IEK-6), Forschungszentrum Jülich, Wilhelm-Johnen-Straβe, 52425 Jülich, Germany
[2] JARA High-Performance Computing, Schinkelstraβe 2, 52062 Aachen, Germany
[3] School of Nuclear Science and Technology, Lanzhou University, Tianshui South Road 222, Lanzhou 730000, China
[4] The Rossendorf Beamline at ESRF– The European Synchrotron, CS40220 38043, Grenoble Cedex 9, France
[5] Helmholtz Zentrum Dresden-Rossendorf (HZDR), Institute of Resource Ecology, PO Box 510119, 01314 Dresden, Germany
[6] School of Chemistry, University of Dublin, Trinity College, College Green, Dublin 2, Ireland
* Corresponding author: e-mail: p.kowalski@fz-juelich.de

**Computational modeling is an important aspect of the research on nuclear waste materials. In particular, atomistic simulations, when used complementary to experimental efforts, contribute to the scientific basis of safety case for nuclear waste repositories. Here we discuss the state-of-the-art and perspectives of atomistic modeling for nuclear waste management on a few cases of successful synergy of atomistic simulations and experiments. In particular, we discuss here: (1) the potential of atomistic simulations to investigate the uranium oxidation state in mixed-valence uranium oxides and (2) the ability of cementitious barrier materials to retain radionuclides such as $^{226}$Ra and $^{90}$Sr, and of studtite/metastudtite secondary peroxide phases to incorporate actinides such as Np and Am. The new contribution we make here is the computation of the incorporation of Sr by C-S-H (calcium silicate hydrate) phases.**

## INTRODUCTION

Formulation of a good scientific basis of safety case for nuclear waste repositories requires a highly interdisciplinary approach and expertise coming from various research fields. These include chemistry, physics, mineralogy, geology, hydrogeology and simulation and data science, to name but a few. In our approach we focus on materials science aspects and conduct in particular research on solid waste forms, including spent nuclear fuel itself, engineered barriers materials or secondary phases that are expected to form under repository conditions. We are interested in the characterization of these materials to assess their capabilities for radionuclide uptake and hosting, durability and long-time performance. Solid state chemistry of radionuclides is an important aspect of such research as it helps to understand the chemical durability of the disposed nuclear waste [1]. With the tremendous progress in efficiency of computing power and performance of software, including quantum chemistry methods such as DFT (Density Functional Theory) [2], over the last two decades computer simulations have been extensively used in the research on nuclear materials [3,2]. In this contribution we focus on the development and selected applications of such methods we made at IEK-6 Institute at Forschungszentrum Jülich in the last years and the lessons learned from joint atomistic simulations and experimental studies of selected problems related to the disposal of nuclear wastes.

Among different methods of computational chemistry and materials science, DFT became the workhorse of the atomistic modeling efforts. This comes as a compromise between accuracy of the method and its computational feasibility. DFT requires the integration of three dimensional electrons charge density and is thus much faster than a direct integration of the Schrödinger equation that must be performed in multi-dimensional space or than any approximate wave functions-based methods (e.g. Hartree-Fock-based methods). Although it is still computationally intensive, DFT permits simulations of systems consisting of a few hundred atoms, including complex solids, fluids and melts [2,4,5]. So far, DFT is the only method allowing calculations of materials properties such as phonon dispersion, heat capacities or direct ab initio molecular dynamics simulations, and we foresee that this will hold for decades.

In principle, DFT is an exact computational method, equivalent to solving the Schrödinger equation. The only source of uncertainty is the unknown exact form of the so-called exchange-correlation functional. The majority of approximations originate from the exact solution for homogeneous electron gas that well describes smooth electronic density and light elements (*s* and *p* orbitals) [6]. In the research on nuclear waste we mainly deal with intermediate mass d elements, 4f lanthanides and heavy *5f* actinides. However, *d* and *f* electrons are strongly correlated and materials that contain them are often poorly described by DFT. As a striking example, actinide dioxides ($AnO_2$) - simple wide band gap insulators - are described by DFT as metals [7]. The DFT reaction enthalpies are also of significant error (up to ~200 kJ/mol for An-bearing compounds) [8-12]. To overcome these problems, an extension of the DFT method, DFT+*U*, that explicitly accounts for electronic correlation, is often applied. This is done by adding the so-called Hubbard term to the Hamiltonian with a parameter *U* (the Hubbard parameter), which purpose is to describe the strong, on-site coulomb repulsion between *d* or *f* electrons [13]. The DFT+*U* method has been successfully applied by various research groups to computation of strongly correlated electrons bearing systems [7,14-16], but in most of these studies the Hubbard parameter *U* is treated as a free parameter. To preserve the pure ab initio character of the method and to obtain the best possible solutions for systems of interest we compute this parameter from first principles (e.g. [10,18,19]).

In this contribution we review our recent activities on application of the ab initio computational approaches to radionuclide bearing systems. We will discuss the benefits of atomistic modeling methods to understand the structural incorporation of radionuclides into different phases, the related oxidation state chemistry and thermodynamics of the process. In particular, we will show how the synergy of atomistic simulations and experimental methods leads to more complete characterization of materials properties.

## COMPUTATIONAL METHODS

The DFT-based ab initio calculations were performed with Quantum-ESPRESSO package [20], by applying different exchange-correlation functionals, including PBE [21] and PBEsol [22]. The core electrons were represented by ultrasoft pseudopotentials [23] and the plane-wave energy cutoff was set to 50 Ryd. The Hubbard $U$ parameter values were derived using the linear response method of Cococcioni & de Gironcoli [13]. The hydration enthalpies and entropies needed for computation of the free energies of reactions were taken from the available databases [24,25]. The entropies of solid phases were estimated using the Latimer approach [26]. The details on the specific calculations are reported in the original publications, which are indicated through the text.

## RESULTS & DISCUSSION

### PERFORMANCE OF THE DFT+U METHOD

Because standard DFT fails for most materials of interest in nuclear waste management, the DFT+$U$ method has been widely used in computation of actinide-bearing nuclear materials. In most of the previous studies on uranium compounds, the applied Hubbard $U$ parameter is either fixed to 4.5 eV (with parameter $J$=0.52 eV describing on-site Coulomb exchange) or selected so that the calculations reproduce certain materials properties, such as the band gap or lattice parameters [7,14,15]. The former, fixed estimate comes from measurements of the correlation energy of $UO_2$ [27]. However, our first principle calculations have shown that the Hubbard $U$ parameter can be significantly different for structurally different compounds and for uranium in different oxidation state [19,28]. There is a clear trend indicating that the Hubbard $U$ parameter value strongly depends on the oxidation state of uranium. It is of ~3 eV for U(VI) and ~2 eV for U(IV) compounds [10,11,19]. The variation of the Hubbard $U$ parameter for uranium oxides is illustrated in Fig. 1. Here, a linear-like increase of the $U$ parameter with the average oxidation state is clearly visible. We note that the obtained values are lower than the commonly used value of 4.5 eV. Very similar results were obtained when calculating other actinides (Pu, Am and Cm [11]) and Ln-bearing phosphate systems [28], including calculations with the cRPA (Constrained Random Phase Approximation) method [29].

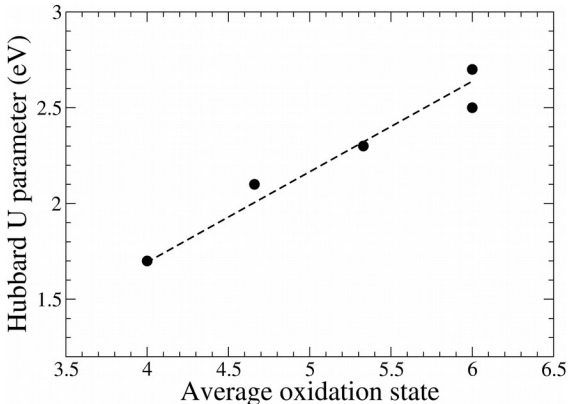

**Figure 1**. The computed Hubbard $U$ parameters for series of uranium oxides. Data from Beridze & Kowalski [8] and Kvashnina et al. [19]. The dashed line is shown to visualize the trend.

A very important conclusion coming out of these studies is that the standard DFT approach with $f$ electrons computed ab inito often results in biased predictions of materials properties. In order to obtain improved description of the f-electrons-bearing systems, the DFT+$U$ or so-called *f in the core* approaches have to be used. In the *f in the core* method the $f$ electrons are not explicitely computed, but their presence is mimicked by the pseudopotential (this method works well for more localized *4f* electrons). Ideally, the first approach could be used to obtain good structural data, while the second works well for fast and well converging calculations of the energy- and elasticity- related properties [30-32]. On the other hand, we noticed that regarding the energies, the standard DFT method with *f in the core* method gives results that are consistent with the DFT+$U$ approach, although it gives much worse structural parameters [28]. We also noticed that the *f in the core* method results in very good description of the heat capacities (thus phonons) [30,31], the excess properties of mixing of solid solutions and the elastic properties [32,33].

## HERFD AND RIXS METHODS

X-ray absorption near edge spectroscopy (XANES) is a widely used technique to reveal the local and electronic structure of matter. When the X-rays hit the sample, an electron is promoted from the ground state level to the first unoccupied state (Figure 2). However, the core hole that is created by that process is very unstable and very quickly filled by an electron from the other levels. The X-ray photons emitted during that process are measured by X-ray emission spectroscopy (XES). Therefore, both techniques provide complementary information about the occupied and unoccupied states. Standard XES is recorded with non-resonant excitations. The XES process recorded with resonant excitations is known as resonant X-ray emission spectroscopy (RXES) and resonant inelastic X-ray scattering (RIXS).

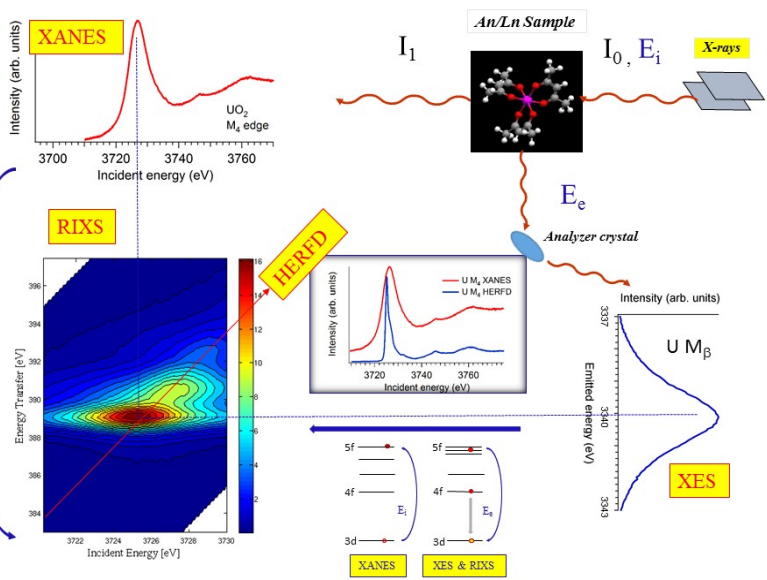

**Figure 2**. Schematic representation of the XANES, HERFD, XES and RIXS measurements at the U $M_4$ edge for $UO_2$ and electronic transitions of those processes. RXIS data are shown as contour maps in a plane of incident ($E_i$) and transferred photon energies ($E_t$), where the vertical axis represents the energy difference between the incident and the emitted energies ($E_e$). Variations of the color in the plot relate to different scattering intensities. The HERFD spectrum corresponds to a diagonal cut through the RIXS planes at the maximum of the Mβ emission line.

The main edge of the U M4 XANES spectrum arises from the electronic transitions from the U $3d_{3/2}$ to $5f$ level [34,35]. The U M4 edge absorption features can be recorded in high energy resolution fluorescence detection (HERFD) mode with the help of the crystal analyzer, installed in the X-ray emission setup [36]. The emission spectrometer is tuned to the $M_{Lx}$ ($4f_{5/2}$-$3d_{3/2}$) transition and the XANES is recorded by monitoring the $M_{Lx}$ intensity as a function of the incident energy. The advantage of such a setup is twofold: the width of the spectral features is no longer limited by the $3d_{3/2}$ core hole lifetime but the sharper $4f_{5/2}$ core hole width in the final state.

The typical procedure to perform HERFD studies is to record the RIXS map near the selected emission energy and to perform analysis of the results using the atomistic modeling support. The HERFD spectrum corresponds to a diagonal cut through the RIXS planes at the maximum of the $M_{Lx}$ emission line (c.f. Fig.2). Special attention has to be focused on the features that do not lie on the diagonal cut (related to HERFD). If all the features are situated along the diagonal direction, further measurements and analysis can be performed by a line scan – referred to the HERFD. Such types of the RIXS are named core-to-core RIXS. However, there are the transitions between the ground state and the valence lines that can be detected by RIXS (named core-to-valence). The intermediate state in core-to-core and core-to-valence RIXS is the same and exhibits a core hole, making the technique element-selective. The energy transfer in core-to-core RIXS near the U $M_{L\bar{x}}$ emission line is very large (~380 eV) and contains a core hole. In core-to-valence RIXS, the decay directly involves valence electrons. The energy transfer is only a few eV, and no core hole is present in the final state.

In this manuscript we report results obtained with two methods: HERFD at the U M4 edge for studies of peroxide phases and valence band RIXS at the U M4 edge method in studies of mixed uranium oxides.

## SIMULATION OF U-OXIDE MATERIALS

RIXS and HERFD methods give an excellent opportunity to investigate the electronic structure and the redox chemistry of mixed uranium oxides [34,35,37]. This also allow for validation of the computational methods, like DFT+$U$. In recent studies of Kvashnina et al. [19] we tested performance of the DFT+$U$ method, with the Hubbard $U$ parameter derived ab initio, for description of various mixed-valence uranium oxides. The best description of RIXS spectra has been obtained with the Hubbard $U$ parameter derived from the linear response method [13] and with the maximally localized Wannier functions as representation of $f$ orbitals. The result for the $U_3O_8$ system is given in Fig. 3. For this case, we obtained the Hubbard $U$ parameter values for U(VI) and U(V) - the oxidation states present in the system – as 2.2 eV and 2.0 eV, respectively. Actually, the valence band RIXS data include the elastic (at 0 eV in Fig. 3) and inelastic scattering profiles (in the range of 4-8 eV in Fig. 3) with an energy resolution of ~1eV and provide information on the energy difference between the valence band states and the unoccupied U $5f$ states – that's why the validity of the electronic structure calculations can be tested. The measured RIXS profiles are reproduced very well by the computed spectrum. Additionally, we were able to compute the U oxidation states in different mixed-valence oxides, which are in agreement with experimentally obtained oxidation states [34,37]. The results are reported in Table 1

**Table 1**. Oxidation states of different U oxide compounds. Data from Kvashnina et al. [19].

| Uranium oxide | Uranium oxidation states |
|:---:|:---:|
| $UO_2$ | U(IV) |
| $U_3O_7$ | U(IV),U(V) |
| $U_3O_8$ | U(V),U(VI) |
| $UO_3$ | U(VI) |

Studtite and metastudtite peroxide phases could be formed under disposal conditions in the case of $UO_2$-based spent nuclear fuel exposed to water [38]. We used atomistic modeling to interpret the HERFD spectra and validate the structural models of the two phases, including the local U environments [38]. In recent follow-up studies, we computed incorporation energies of Np and Am into both phases [39]. The most energetically favorable oxidation states and their solution energies are reported in Table 2. We found that the incorporation of both elements is thermochemically endothermic, which explains the inability to incorporate larger amounts of these elements into the peroxide phases. However, the computed values show much better probability of incorporating Np that Am, with difference in incorporation energies of ~0.5 to 1 eV.

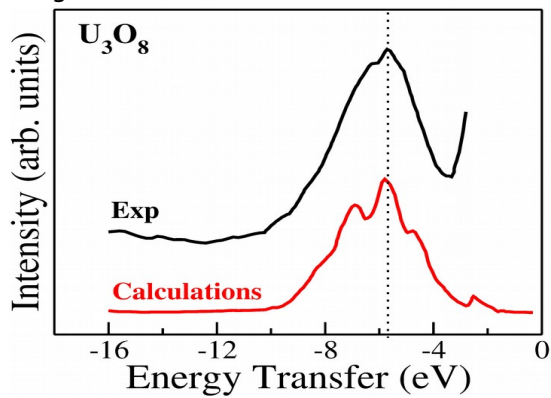

**Figure 3**. The measured and computed RIXS for $U_3O_8$ system. Data are taken from Kvashnina et al. [19].

**Table 2**. The Np and Am incorporation energies (in eV) into studtite and metastudtite peroxides. Data from Biswas et al. [39].

|        | studtite | metastudtite |
|--------|----------|--------------|
| Np(VI) | 1.12     | 1.08         |
| Np(V)  | 1.43     | 1.35         |
| Am(IV) | 1.95     | 1.64         |
| Am(III)| 2.01     | 1.47         |

## RETENTION OF Ra AND Sr BY CEMENTITIOUS MATERIALS

Cementitious materials are widely used in nuclear waste management, including repository concepts. These are utilized as back-fill in repositories, for solid wastes encapsulation or as component of waste containers. One of the most important topics is the retention of radionuclides by these phases. We performed an experimental investigations of the phases present in hardened cement paste to investigate the uptake of Ra by single cement hydration phases [40]. The selected results are given in Fig. 4. There is a significant deviation in the Ra uptake capability with C-S-H phases being able to uptake significant amounts of Ra from aqueous solution. In order to understand the structural incorporation of Ra by these compounds and the thermodynamics of the process we performed careful ab initio studies of a series of C-S-H phases with varying Ca/Si ratio (0.75<Ca/Si<1.0). These phases have been computed using the 11 Å tobermorite model. It has two Ca cations sublattices: a fully saturated intralayer and a partially saturated interlayer that contain water. The calculations show a significant energy difference for cation exchange (Ca → Ra) in intra and interlayer, with the interlayer being preferred by as much as ~200 kJ/mol. This conclusively shows that Ra can be incorporated in the interlayer only. We found that the incorporation enthalpy and free energy, the later estimated form the available thermodynamic data and theoretical entropies of elements in solid phases [24-26], increase with increasing Ca content, reaching values of -18.5 kJ/mol and 6 kJ/mol, respectively, for the Ca/Si ratio of 0.9. This is consistent with the experimentally seen trend of the reduced uptake capability with the increase of Ca content.

In this contribution we complement the studies of Lange et al. [40] on the uptake of Ra with calculations regarding the incorporation of Sr. Previous studies of Tits et al. [41,42] indicate significant difference in the uptake of Ra and Sr by C-S-H, with uptake of Sr being two orders of magnitude lower. The measured Kc values (selectivity constants) for the cation exchange reactions are shown in Fig. 5. In order to understand this difference we have computed enthalpies and free energies of the relevant cation exchange reaction (see Lange et al. [40] for details on the computational procedure). The results for C-S-H with Ca/Si=0.9 are reported in Table 3. The values for Sr are larger. The difference in the free energy between the cases of Ra and Sr of 12.7 kJ/mol should result at ambient conditions in a factor of ~160 difference in uptake capability (applying Boltzmann exponential factor), which is indeed seen experimentally (Kc values, Fig. 5). The main contributor to this difference is the significant difference in the reaction entropy, which is ~-99.1 J/mol/K for Sr and ~-83.3 J/mol/K for Ra cases. We also found that the largest uncertainties come from uncertainties in the thermodynamic data that result in an error of ~20 kJ/mol in the cation exchange reaction free energy.

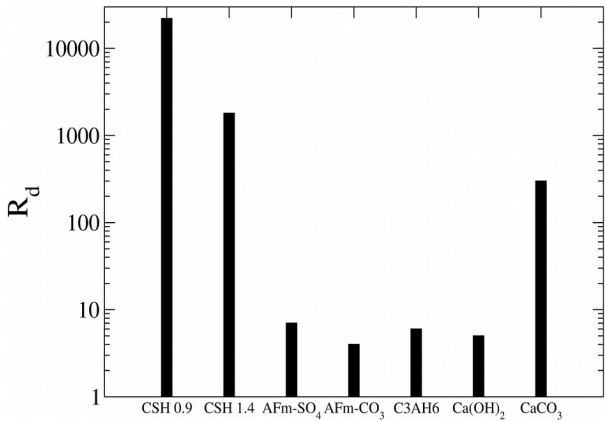

**Figure 4**. The uptake of Ra ($R_d$ values) by selected cement hydration phases as determined by Lange et al. [40].

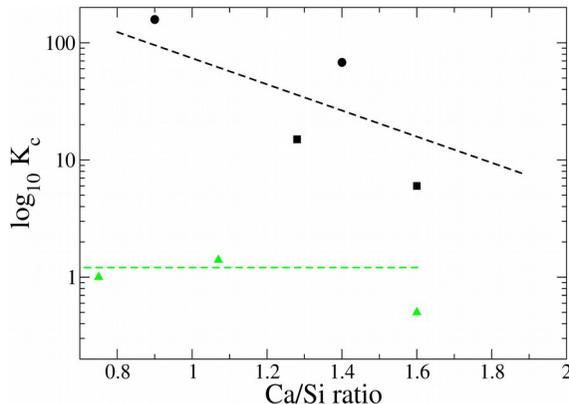

**Figure 5**. The measured $K_c$ values for the Ra and Sr uptake by C-S-H phases as a function of Ca/Si ratio. Data from Lange et al. [40] (Ra: black circles) and Tits et al. [41,42] (Ra: black squares; Sr: green triangles). The dashed lines are shown to visualize the trends.

**Table 3**. The computed cation exchange reaction enthapies and free energies for incorporation of Ra and Sr into C-S-H phase with Ca/Si ratio =0.9. Data for Ra from Lange et al. [40].

| Cation | ΔH (kJ/mol) | ΔH (kJ/mol) |
|---|---|---|
| Ra | -18.5 | 6.5 |
| Sr | -10.5 | 19.2 |

## CONCLUSIONS

We discussed the state-of-the-art and current perspectives of the application of atomistic modeling on selection of a few examples of the recently performed own research. Our experience shows that strong electronic correlations have to be carefully accounted for in order to correctly capture the electronic structure and thermochemistry of the considered systems. We discussed the successful application of our methodologies in interpretation of HERFD and RIXS for mixed-valence uranium oxide and uranium peroxide phases. In the first case we deliver conclusive results regarding the oxidation state of uranium. In the latter case, we derived Np and Am solubility energies in these phases, explaining the difficulty to incorporate these actinides into the peroxide phases, especially Am. By computing the incorporation of Ra into C-S-H phases we demonstrated the feasibility of atomistic modeling for understanding the incorporation mechanism and thermodynamics of the process. The performed here supplemental calculation of Sr shows the reasoning for the much smaller uptake of Sr by C-S-H phases compared to Ra.

The presented examples show that the recent improvement in the atomistic modeling and the synergy of experimental and modeling studies can result in superior characterization of materials. Nevertheless, computation of *d* and *f* elements is computationally challenging but with further development of supercomputing resources, methods and increase of available experimental data sets we foresee further improvements and even more successful joint atomistic modeling and experimental studies in the future.

**ACKNOWLEDGMENTS**

The work was funded by the Excellence Initiative of the German federal and state governments and the Jülich Aachen Research Alliance in High-Performance Computing. We thank the JARA-HPC awarding body for time on the JURECA and RWTH computing cluster awarded through JARA-HPC Partition. The research leading to these results has received funding from the European Union's Horizon 2020 Research and Training Programme of the European Atomic Energy Community (EURATOM) (H2020-NFRP-2014/2015) under grant agreement n° 662147 (CEBAMA). MS acknowledges the funding from China Scholarship Council (CSC) for study and research in Germany. KOK acknowledges support from European Commission Council under ERC grant No. 759696.

**REFERENCES**

1. R. C. Ewing, PNAS, 96, 3432 (1999).
2. S. Jahn and P. M. Kowalski, Rev. Mineral. Geochem., 78, 691 (2014).
3. A. Chroneos, M. Rushton, C. Jiang and L. Tsoukalas, J. Nucl. Mater., 441, 29 (2013).
4. P. M. Kowalski, B. Wunder and S. Jahn, Geochim. Cosmochim. Acta, 101, 285 (2013).
5. S. Blouin, P. M. Kowalski and P. Dufour, Astrophys. J, 848, 36 (2017).
6. A. D. Becke, Phys. Rev. A, 38, 3098 (1988).
7. X.-D. Wen, R. L. Martin, T. M. Henderson and G. E. Scuseria, Chem. Rev., 113, 1063 (2013).
8. G. A. Shamov, G. Schreckenbach and T. N. Vo, Chem. Eur. J., 13, 4932 (2007).
9. N. Iché-Tarrat and C. J. Marsden, J. Phys. Chem. A, 112, 7632 (2008).
10. G. Beridze and P. M. Kowalski, J. Phys. Chem. A, 118, 11797 (2014).
11. G. Beridze, A. Birnie, S. Koniski, Y. Ji and P. M. Kowalski, Prog. Nucl. Energy, 92, 142 (2016).
12. P. M. Kowalski, G. Beridze, Y. Ji. and Y. Li, MRS Advances, 2, 491 (invited) (2017).
13. M. Cococcioni and S. de Gironcoli, Phys. Rev. B, 71, 035105 (2005).
14. S. O. Odoh, Q. -J. Pan, G. A. Shamov, F. Wang, M. Fayek, G. Schreckenbach, Chem. Eur. J., 18, 7117 (2012).
15. Z. Rák, R. C. Ewing, U. Becker, Phys. Rev. B, 84, 155128 (2011).
16. A. Floris, S. de Gironcoli, E. K. U. Gross and M. Cococcioni, Phys. Rev. B, 84, 161102 (2011).
17. S. Wu, P. M. Kowalski, N. Yu, T. Malcherek, W. Depmeier, D. Bosbach, S. Wang, E. V. Suleimanov, T. E. Albrecht-Schmitt and E. V. Alekseev, Inorg. Chem., 53, 7650 (2014).
18. G. L. Murphy, B. J. Kennedy, J. A. Kimpton, Q. Gu, B. Johannessen, G. Beridze, P. M. Kowalski, D. Bosbach, M. Avdeev, and Z. Zhang, Inorg. Chem., 55, 9329 (2016).
19. K. O. Kvashnina, P. M. Kowalski, S. M. Butorin, G. Leinders, J. Pakarinen, R. Bès, H. Li, and M. Verwerft, Chem. Commun, 54, 9757 (2018).
20. P. Giannozzi et al., J. Phys. Condens. Matter, 21, 395502 (2009), http://www.quantum-espresso.org (accessed on 25.11.2016).
21. J. P. Perdew, K. Burke and M. Ernzerhof, Phys. Rev. Lett., 77, 3865 (1996).
22. J. P. Perdew, A. Ruzsinszky, G. I. Csonka, O. A. Vydrov, G. E. Scuseria, L. A. Constantin, X. Zhou and K. Burke, Phys. Rev. Lett., 100, 136406 (2008).

23. D. Vanderbilt, Phys. Rev. B, 41, 7892 (1990).
24. E. L. Shock, D. C. Sassani, M. Willis and D. A. Sverjensky, Geochem. Cosmochim. Acta, 61, 907 (1997).
25. D. W. Smith, J. Chem. Educ., 54, 540 (1977).
26. W. M. Latimer, J. Am. Chem. Soc., 73 1480 (1951).
27. Y. Baer and J. Schoenes, Solid State Commun., 33, 885 (1980).
28. A. Blanca-Romero, P. M. Kowalski, G. Beridze, H. Schlenz and D. Bosbach, J. Comput. Chem., 35, 1339 (2014).
29. P. M. Kowalski, G. Beridze, Y. Li, Y. Ji, C. Friedrich, E. Sasioglu and S. Blügel, Ceram. Trans., 258, 205 (2016).
30. P. M. Kowalski, G. Beridze, V. L. Vinograd and D. Bosbach, J. Nucl. Mater., 464, 147 (2015).
31. Y. Ji, G. Beridze, D. Bosbach and P. M. Kowalski, J. Nucl. Mater., 494, 172 (2017).
32. P. M. Kowalski and Y. Li, J. Eur. Ceram. Soc., 36, 2093 (2016).
33. Y. Li, P. M. Kowalski, A. Blanca-Romero, V. L. Vinograd and D. Bosbach, J. Solid State Chem., 220, 137 (2014).
34. K.O. Kvashnina, S.M. Butorin, P. Martin, P. Glatzel, Phys. Rev. Lett. 111, 253002 (2013).
35. K.O. Kvashnina, Y.O. Kvashnin, S.M. Butorin, J. Electron Spectros. Relat. Phenomena. 194, 27-36 (2014).
36. K.O. Kvashnina, A.C. Scheinost, J. Synchrotron Radiat. 23, 836–841 (2016).
37. G. Leinders, R. Bes, J. Pakarinen, K. Kvashnina, M. Verwerft, Inorg. Chem. 56, 6784–6787 (2017).
38. T. Vitova, I. Pidchenko, S. Biswas, G. Beridze, P. W. Dunne, D. Schild,, Z. Wang, P. W. Kowalski and R. J. Baker, Inorg. Chem., 57, 1735 (2018).
39. S. Biswas, S. J. Edwards, Z. Wang, H. Si, L. L. Vintró, B. Twamley, P. M. Kowalski and Robert J. Baker, Dalton Trans., 48, 13057 (2019).
40. S. Lange, P. M. Kowalski, M. Pšenička, M. Klinkenberg, S. Rohmen, D. Bosbach and G. Deissmann, Appl. Geochem., 96, 204 (2018).
41. J. Tits, E. Wieland, C. J. Müller, C. Landesman and M. H. Bradbury, J. Colloid Interface Sci., 300, 78 (2006).
42. J. Tits, K. Iijima, G. Kamei and E. Wieland, Radiochim. Acta, 94, 637 (2006).